\documentclass[letterpaper,preprintnumbers,showpacs,english,floats,floatfix,amssymb,prd,twocolumn,nofootinbib]{revtex4-1}
\usepackage{amssymb,amsmath}
\usepackage{epsfig,hyperref}
\usepackage[svgnames]{xcolor}
\usepackage{pgf,tikz}
\usepackage{epstopdf}
\usepackage[british]{babel}
\usepackage{enumerate}
\usepackage{enumitem}
\usepackage{makecell}
\usepackage{booktabs}

\makeatletter
\DeclareRobustCommand*{\bfseries}{%
  \not@math@alphabet\bfseries\mathbf
  \fontseries\bfdefault\selectfont
  \boldmath
}
\makeatother

\def\be{\begin{equation}}
\def\ee{\end{equation}}
\def\beq{\begin{eqnarray}}
\def\eeq{\end{eqnarray}}

\makeatletter

\newcommand{\arXiv}[2][]{\href{http://arxiv.org/abs/#2}{\texttt{arXiv:#2\@ifempty{#1}{}{ [#1]}}}}
\makeatother

\begin{document}
\title{Dynamics of critical collapse}

\author{Jun-Qi Guo}%
\email{sps{\_}guojq@ujn.edu.cn}
\affiliation{School of Physics and Technology,
University of Jinan, No. 336, West Road of Nan Xinzhuang, Jinan 250022, Shandong, China}

\author{Hongsheng Zhang}%
\email{sps{\_}zhanghs@ujn.edu.cn}
\affiliation{School of Physics and Technology,
University of Jinan, No. 336, West Road of Nan Xinzhuang, Jinan 250022, Shandong, China}

\date{\today}

\begin{abstract}
  Critical collapse of a massless scalar field in spherical symmetry is systematically studied. We combine numerical simulations and asymptotic analysis, and synthesize critical collapse, spacetime singularities, and complex science. First set of approximate analytic expressions near the center are obtained. We observe that, near the center, the spacetime is nearly conformally flat, the dynamics is not described by the Kasner solution, and the Kreschmann scalar is proportional to $r^{-5.30}$, where $r$ is the areal radius. These features are significantly different from those in black hole singularities. It is speculated that the scalar field in critical collapse may be a special standing wave.
\end{abstract}
\maketitle

\section{Introduction\label{sec:introduction}}

Complex systems widely exist in Nature. The interactions between the parts, alternatively the nonlinearities, in a complex system make the system more than the sum of its parts, causing the emergence of the intriguing critical phenomena and discrete scale invariance in some circumstances~\cite{Sornette_2006,Sornette_1998}. The critical phenomena in gravitational collapse caused by the nonlinearities of Einstein equations were originally discovered by Choptuik via numerical simulations~\cite{Choptuik:1992jv}. A naked singularity is formed in critical collapse~\cite{Wald:1997wa}. So critical collapse connects two basic fields in science: critical phenomena and spacetime singularities.

Choptuik simulated critical collapse of a massless scalar field in spherical symmetry in general relativity. When the scalar field is weak, the scalar field will implode and then disperse, leaving a flat spacetime behind; while when the scalar field is strong enough, it will implode and form a black hole (BH). By fine tuning the strength of the scalar field, a critical solution which is on the threshold of BH formation can be obtained. The results for critical collapse include discrete self-similarity (DSS), universality, and mass scaling law. The DSS states that the scalar field oscillates periodically at ever-decreasing time and length scales related by a factor of $e^{\Delta}$ with $\Delta\approx3.44$. Regarding the universality, all the families of the near-critical evolutions approach the same solution. The mass scaling law describes that, in the super-critical case, the mass of the tiny BH has a scaling relation with the parameter of the scalar field, $M_{\scriptsize{\mbox{BH}}}\simeq B|p-p^{*}|^{\gamma}$, where $B$ is a family-dependent parameter, $p$ is one parameter for the scalar field, describing the strength of the scalar field, $p^{*}$ is the critical value for $p$, and $\gamma$ is a universal scaling exponent $\gamma\approx0.37$.

After Choptuik's discovery, simulations of critical collapse in other gravitational theories, symmetries, and matter fields have been implemented~\cite{Liebling:1996dx,Sorkin:2005vz,Golod:2012yt,Deppe:2012wk,Choptuik:2003ac,Abrahams:1993,Abrahams:1994,Choptuik:1996yg,Baumgarte:2016xjw}.
Critical collapse of perfect fluids with continuous self-similarity was explored in Refs.~\cite{Evans:1994pj,Maison:1995cc,Koike:1995jm,Hara:1996mc,Koike:1999,Neilsen:1998qc}. It was proposed that the scaling exponent $\gamma$ can be obtained via a linear stability analysis of the critical solution in Ref.~\cite{Evans:1994pj}, and such a proposal was implemented successfully in Refs.~\cite{Maison:1995cc,Koike:1995jm,Hara:1996mc,Koike:1999}. The scaling exponent was also obtained by evolving near-critical initial data numerically in Ref.~\cite{Neilsen:1998qc}, confirming the results from the linear perturbation theory~\cite{Maison:1995cc}.

Considering that the results on critical collapse by Choptuik are numerical, it is natural to study this issue analytically. The existence of a real analytic solution to critical collapse was proved in Ref.~\cite{Reiterer:2012mz}. In Refs.~\cite{Gundlach:1995kd,Gundlach:1996eg,MartinGarcia:2003gz}, under the requirements of discrete scale invariance, analyticity, and an additional reflection-type symmetry, the critical collapse is reduced to an eigenvalue problem. The rescaling factor $\Delta$ becomes an eigenvalue and is solved numerically by a relaxation algorithm with high precision. Motivated by understanding the nature of the echoing of the scalar field, Price and Pullin found that although the oscillatory behavior of the scalar field seems to come from the nonlinearities of general relativity, it can be approximated by a scalar field solution in flat spacetime~\cite{Price:1996sk}. In addition, analytic models for the continuous self-similar collapse in spherical symmetry (Roberts solution)~\cite{Roberts,Oshiro:1994hd,Brady:1993np} and cylindrical symmetry~\cite{Wang:2003vf} were presented. In Ref.~\cite{Frolov:1999fv}, with analytic perturbation methods, it was found that a generic perturbation departs from the Roberts solution in a universal way. For reviews on critical collapse, see Refs.~\cite{Choptuik:1997mq,Gundlach:2007gc,Choptuik:2015mma}.

Despite the above efforts on analytic studies, due to the complexity of Einstein equations, ever since the publication of Choptuik's numerical results in 1993, the analytic solution remains unknown, and the nature of critical collapse is far from being well understood. In this paper, we explore the dynamics of critical collapse by combining numerical simulations and asymptotic analysis. Considering that critical collapse ends up with a naked singularity, we connect critical collapse to the results on the dynamics near spacetime singularities that have been obtained before, including the Belinskii, Khalatnikov, and Lifshitz (BKL) conjecture and BH formation. With such efforts, some approximate analytic solutions near the center are obtained. It is found that, near the center, the spacetime is nearly conformally flat, and the dynamics is not described by the Kasner solution.

The paper is organized as follows. In Sec.~\ref{sec:framework}, we describe the framework. Section~\ref{sec:conformal_flatness} presents the nearly conformal flatness of the spacetime in critical collapse. In Sec.~\ref{sec:analytic_slns}, approximate analytic information on critical collapse is reported. We compare critical collapse with the BKL conjecture and BH formation in Sec.~\ref{sec:comparison}. In Sec.~\ref{sec:standing_wave}, the connection between the scalar field in critical collapse and standing waves is argued. In Sec.~\ref{sec:discussions}, the results are discussed.
Throughout the paper, we set $G=c=\hbar=1$.

\section{Framework\label{sec:framework}}
In this section, we present the framework for numerical simulations of critical collapse. Critical collapse of a massless scalar field in spherical symmetry is simulated in double-null coordinates,
\be
\begin{split}
ds^{2} &= -4e^{-2\sigma(u,v)}dudv+{r^{2}(u,v)}d\Omega^2\\
&= e^{-2\sigma(t,x)}(-dt^2+dx^2)+{r^{2}(t,x)}d\Omega^2,
\end{split}
\label{double_null_metric}
\ee
where $u=(t-x)/2$ and $v=(t+x)/2$. Consider a massless scalar field $\psi$ with the energy-momentum tensor
$T_{\mu\nu}=\psi_{,\mu}\psi_{,\nu}-(1/2)g_{\mu\nu}g^{\alpha\beta}\psi_{,\alpha}\psi_{,\beta}$. Then the equations of motion can be written as~\cite{Frolov_2004,Guo:2013dha}
\be r(-r_{,tt}+r_{,xx})-r_{,t}^2+r_{,x}^2 = e^{-2\sigma},\label{equation_r}\ee
\be -\sigma_{,tt}+\sigma_{,xx} + \frac{r_{,tt}-r_{,xx}}{r}+4\pi(\psi_{,t}^2-\psi_{,x}^2)=0,\label{equation_sigma}\ee
\be -\psi_{,tt}+\psi_{,xx}+\frac{2}{r}(-r_{,t}\psi_{,t}+r_{,x}\psi_{,x})=0. \label{equation_psi}\ee
The constraint equations are~\cite{Frolov_2004,Guo:2013dha}
\be r_{,tx}+r_{,t}\sigma_{,x}+r_{,x}\sigma_{,t}+4\pi r\psi_{,t}\psi_{,x}=0,
\label{constraint_eq_xt}
\ee
\be
r_{,tt}+r_{,xx}+2r_{,t}\sigma_{,t}+2r_{,x}\sigma_{,x}+4\pi r(\psi_{,t}^2+\psi_{,x}^2)=0.
\label{constraint_eq_xx_tt}
\ee
In this paper, $r_{,t}$ is defined as $r_{,t}\equiv\partial{r(t,x)}/\partial{t}$, and other quantities, e.g., $r_{,x}$, $r_{,tt}$, etc, are defined analogously.

For numerical stability concern, the Misner-Sharp mass $m$ is used as an auxiliary variable as has been successfully implemented in Ref.~\cite{Csizmadia:2009dm},
\be g^{\mu\nu}r_{,\mu}r_{,\nu}=e^{2\sigma}(-r_{,t}^2+r_{,x}^2){\equiv}1-\frac{2m}{r}.\label{mass_definition}\ee
Equation~(\ref{mass_definition}) is one new constraint. Then Eqs.~(\ref{equation_r}) and (\ref{equation_sigma}) can be rewritten as
\be -r_{,tt}+r_{,xx} - e^{-2\sigma}\cdot\frac{2m}{r^2}=0,\label{equation_r_2}\ee
\be -\sigma_{,tt}+\sigma_{,xx} - e^{-2\sigma}\cdot\frac{2m}{r^3}+4\pi(\psi_{,t}^2-\psi_{,x}^2)=0.\label{equation_sigma_2}\ee
The dynamics of $m$ is described by
\be
m_{,t}=4\pi r^2\cdot e^{2\sigma}\left[-\frac{1}{2}r_{,t}(\psi_{,t}^2+\psi_{,x}^2)+r_{,x}\psi_{,t}\psi_{,x}\right].
\label{dmdt}
\ee
The quantity $m_{,x}$ is
\be
m_{,x}=4\pi r^2\cdot e^{2\sigma}\left[\frac{1}{2}r_{,x}(\psi_{,t}^2+\psi_{,x}^2)-r_{,t}\psi_{,t}\psi_{,x}\right].
\label{dmdx}
\ee
The derivation of Eqs.~(\ref{dmdt}) and (\ref{dmdx}) are given in the Appendix. In the simulations, Eqs.~(\ref{equation_psi}), (\ref{equation_r_2}), (\ref{equation_sigma_2}), and (\ref{dmdt}) are integrated numerically with finite difference method and leapfrog integration scheme.

We impose $r_{,tt}=r_{,t}=\sigma_{,t}=\psi_{,t}=0$ at $t=0$. The initial profile of $\psi$ is set as
\be
\psi(t=0,x)=a\cdot\exp\left[-\frac{(x-x_0)^2}{b}\right],
\ee
with $a$ being tuned as $a=0.0908379681$, $b=0.01$, and $x_{0}=0.25$. We set $r=\sigma=m=0$ and $r_{,x}=1$ at the origin $(x=0,t=0)$. The expressions for $\sigma_{,x}$, $r_{,xx}$, and $m_{,x}$ can be obtained from Eqs.~(\ref{constraint_eq_xx_tt}), (\ref{equation_r_2}), and (\ref{dmdx}), respectively. We obtain the quantities $r$, $\sigma$, and $m$ on the initial slice of $t=0$ by numerically integrating such expressions from $x=0$ to $x=2$ with the fourth-order Runge-Kutta method. From Eqs.~(\ref{equation_psi}), (\ref{equation_r_2}), and (\ref{equation_sigma_2}), the values of $\psi$, $r$, and $\sigma$ at $t={\Delta}t$ can be determined via a second-order Taylor expansion. Take the first-order time derivative of Eq.~(\ref{dmdt}), one can obtain $m_{,tt}$. With $m_{,t}$ and $m_{,tt}$, $m$ at $t={\Delta}t$ is obtained via a second-order Taylor series expansion.

Regarding the boundary conditions, we always set $r=m\equiv0$ at $x=0$. Then there are always $r_{,t}={r_{,tt}}\equiv0$ at $x=0$. From Eqs.~(\ref{equation_psi}) and (\ref{equation_r_2}), one obtains respectively $\psi_{,x}\equiv0$ and $r_{,xx}\equiv0$ at $x=0$. Then with Eq.~(\ref{constraint_eq_xx_tt}), there is $r_{,x}\sigma_{,x}=0$ at $x=0$. Since $r_{,x}$ is usually not zero, one obtains $\sigma_{,x}\equiv0$ at $x=0$.

The simulations of critical collapse need to be highly accurate. In order to achieve this objective, adaptive or fixed mesh refinement techniques are usually used. While in this paper, we take an even simpler approach. From the beginning, we use very small spatial and temporal grid
spacings ${\Delta}x={\Delta}t=2.5\times10^{-5}$, not making any mesh refinement throughout the whole simulations. It turns out that the numerical results by this approach can show the basic features of critical collapse and are adequate for us to study the dynamics of critical collapse. The code is second-order convergent.

\begin{figure*}[t!]
  \begin{tabular}{ccc}
  \epsfig{file=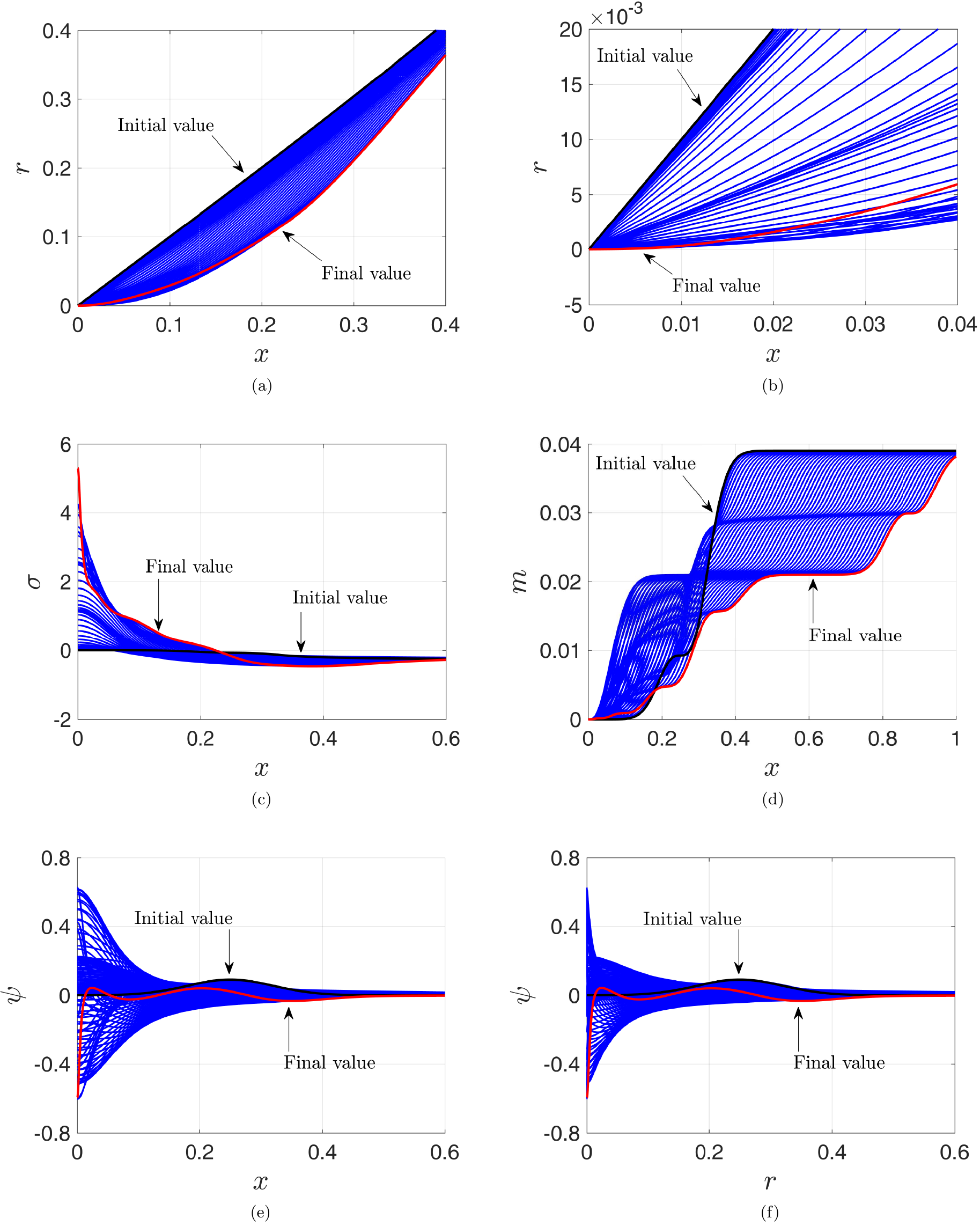, width=0.9\textwidth}
  \end{tabular}
  \caption{(color online). Evolution of the metric components and scalar field on consecutive time slices. The time interval between two consecutive time slices in (a)-(d) is $400\Delta t=0.01$, and in (e) and (f) is $200\Delta t=0.005$. The total time in the simulations is $24360\Delta t=0.609$. (b) Near the center, $r(t,x)\approx D(t)x$. So for fixed $t$, $r$ is a linear function of $x$.}
  \label{fig:evolution}
\end{figure*}

\begin{figure*}[t!]
  \begin{tabular}{ccc}
  \epsfig{file=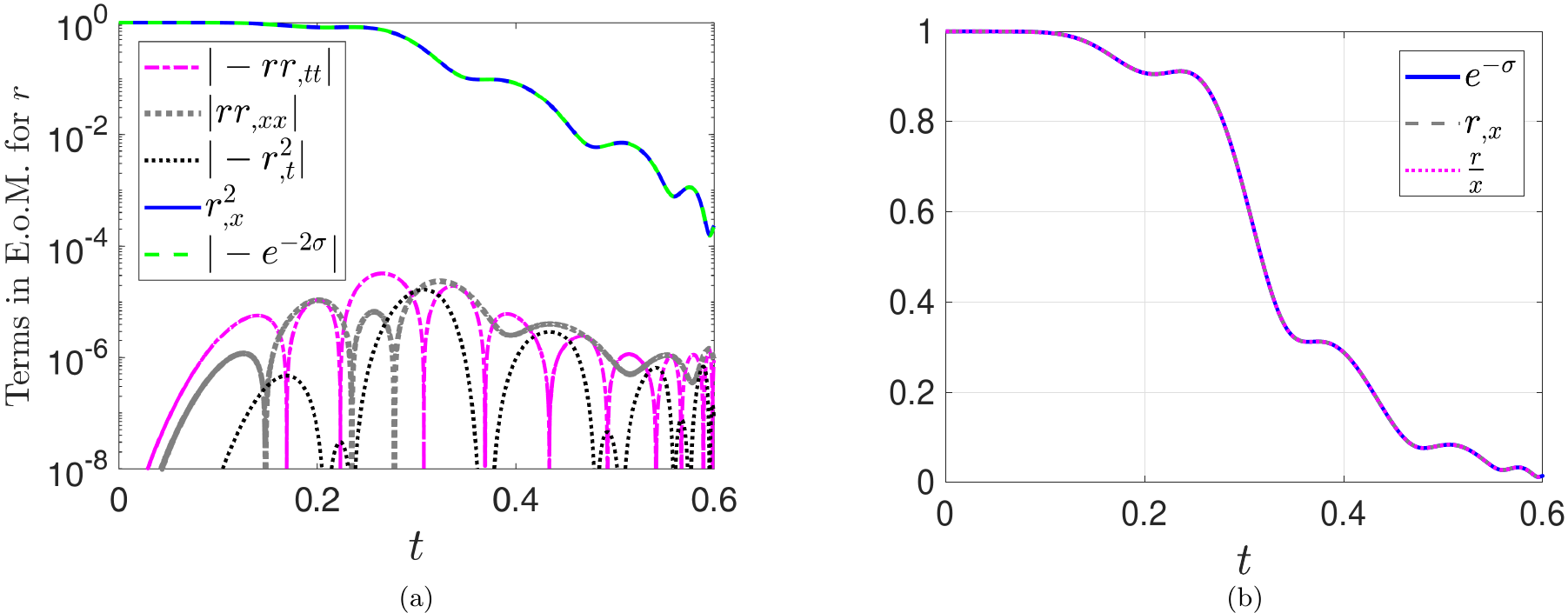, width=0.9\textwidth}
  \end{tabular}
  \caption{(color online). Contributions of all the terms in the equation of motion for $r$~(\ref{equation_r}).
  (a) $e^{-\sigma}\approx r_{,x}$. (b) $r_{,x}\approx r/x$. Slice: $x=4.75\times 10^{-4}$.}
  \label{fig:eom_r}
\end{figure*}

\begin{figure}[t!]
  \begin{tabular}{ccc}
  \epsfig{file=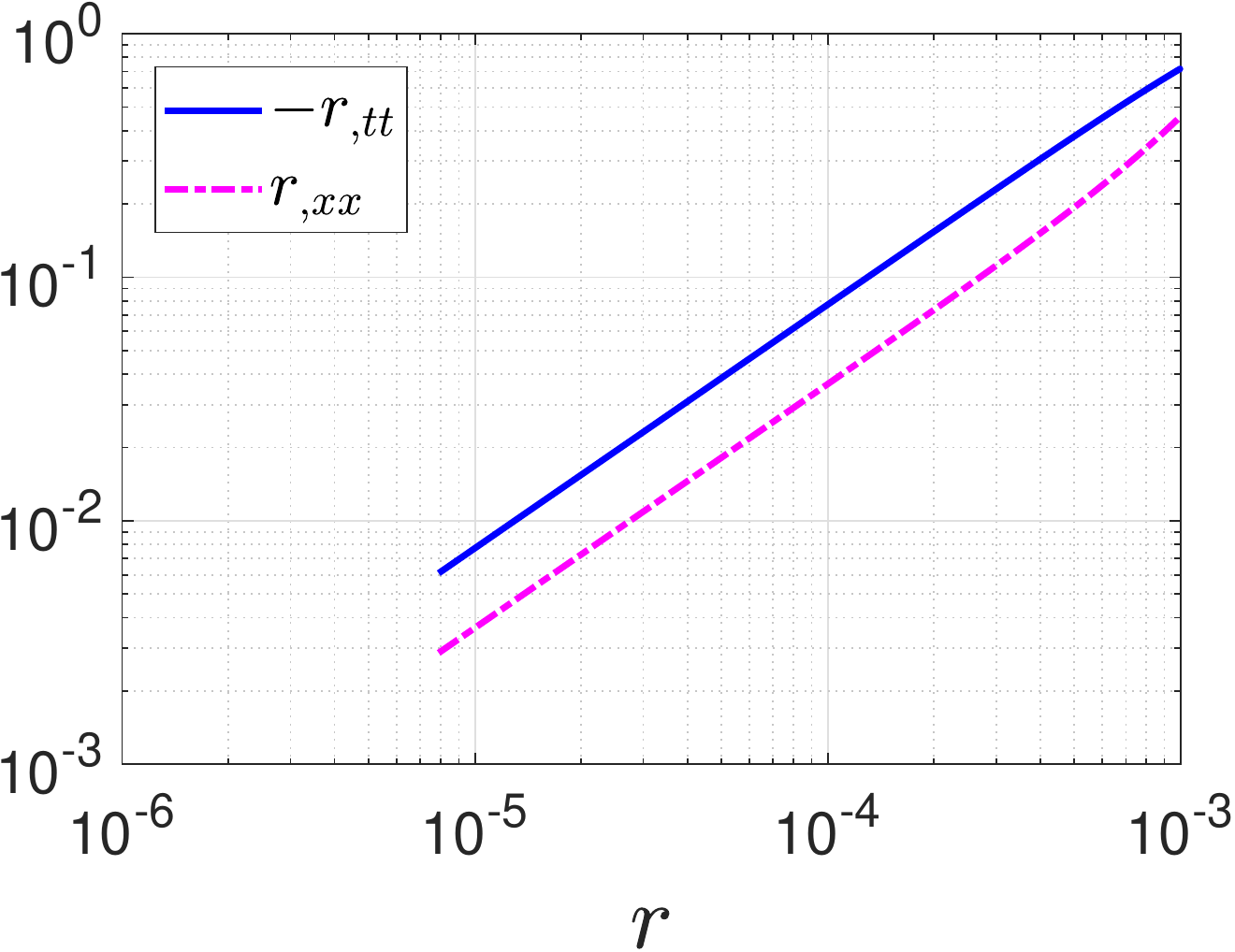,width=7.1cm}
  \end{tabular}
  \caption{(color online). $(-r_{tt},r_{xx})$ vs. $r$ near the center. Slice: $t=0.52$.}
  \label{fig:r_xx}
\end{figure}

\begin{figure*}[t!]
  \begin{tabular}{ccc}
  \epsfig{file=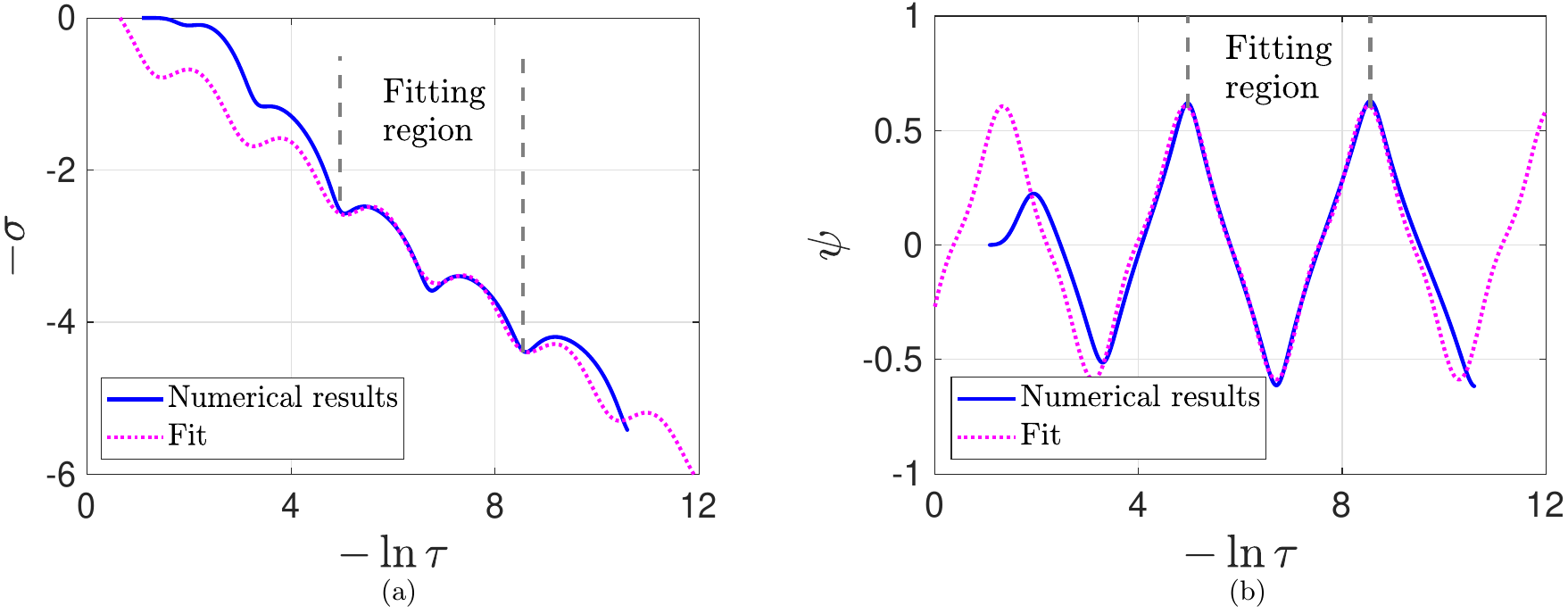, width=0.9\textwidth}
  \end{tabular}
  \caption{(color online). $(-\sigma,\psi)$ vs. $-\ln\tau$. The numerical results are fit for $4.96\le-\ln\tau\le8.55$, correspondingly for $0.4730\le t\le0.5945$. (a) The numerical results on $-\sigma$ vs. $-\ln\tau$ are fit according to $f(\zeta)=a\zeta+\ln[1+b\cos(\omega\zeta+c)]+d$.
  The period for $\sigma$ is $T_{\sigma}\approx 1.795$, and $\omega$ is fixed to be $\omega=2\pi/T_{\sigma}=3.50$. The fitting results are $a=-0.5028\pm0.0015$, $b=0.2287\pm0.0018$, $c=-1.482\pm0.009$, and $d=0.1757\pm0.0096$.
  (b) The numerical results on $\psi$ vs. $-\ln\tau$ are fit according to
  $f(\zeta)= a_{0}+\sum\limits_{n=1}^{4}a_{n}\cos(nw\zeta)+b_{n}\sin(nw\zeta)$. The period for $\psi$ is $T_{\psi}\approx 3.59$, and $\omega$ is fixed to be $\omega=2\pi/T_{\psi}=1.75$.
  The fitting results are $a_{0}=0.0106\pm0.0004$,
  $a_{1}=-0.3451\pm0.0006$,
  $b_{1}=0.3927\pm0.0007$,
  $a_{2}=0.0100\pm0.0006$,
  $b_{2}=-0.0001\pm0.0006$,
  $a_{3}=0.0546\pm0.0006$,
  $b_{3}=0.0508\pm0.0006$,
  $a_{4}=0.0007\pm0.0006$,
  $b_{4}=0.00035\pm0.00059$.
  Slice: $x=4.75\times 10^{-4}$.}
  \label{fig:sigma_psi_vs_tau}
\end{figure*}

\section{The spacetime near the center is nearly conformally flat\label{sec:conformal_flatness}}
The numerical results for the evolution of $r$, $\sigma$, $m$, and $\psi$ are shown in Fig.~\ref{fig:evolution}. Near the center, $\sigma$ goes to $+\infty$. The scalar field $\psi$ oscillates, and at the same time, moves toward the center under gravity. It is noticeable that there are at least two basic parameters for a wave: period and amplitude. However, the amplitude of the scalar field has not been paid as much attention as the period has in the literature in the past. As shown in Figs.~\ref{fig:evolution}(e) and \ref{fig:sigma_psi_vs_tau}(b), the amplitude is about $0.61$. We note that, in Choptuik's original work, where the polar-areal gauge was used, the amplitude is about $0.45$~\cite{Choptuik:2003ac}. In the work by Akbarian and Choptuik~\cite{Akbarian:2015oaa} and the work by Baumgarte~\cite{Baumgarte:2018fev}, where the Baumgarte-Shapiro-Shibata-Nakamura formulation of Einstein's equations in spherical polar coordinates, the 1+log slicing condition for the lapse and the Gamma-driver condition for the shift were implemented, the amplitude is about $0.61$. One may obtain additional hints on the nature of and possible analytic solution to critical collapse by pondering the amplitude and period together.

Noting that the dynamics is simply determined by the equations, we examine the behavior of each term in each equation by plotting the numerical results. We investigate the contribution of each term in the equation of motion for $r$~(\ref{equation_r}) on the slice of $x=4.75\times 10^{-4}$. Remarkably, as shown in Fig.~\ref{fig:eom_r}(a), near the center, ever since the beginning of the collapse, there is always $e^{-\sigma}\approx r_{,x}$. Besides, analytic results means relations between different quantities, so we check the relations between quantities by plotting the corresponding numerical results, e.g., $r_{,x}$ vs. $r$. We plot $r_{,x}$ vs. $r$ on the same slice, and find that $r_{,x}\approx r/x$, see Fig.~\ref{fig:eom_r}(b). Then there is $r(t,x)\approx D(t)x$, where $D(t)$ is a function of $t$. Combine these results,
\be e^{-\sigma}\approx r_{,x}\approx\frac{r}{x}\approx D(t).\label{eom_r_4}\ee
Substitution of Eq.~(\ref{eom_r_4}) into Eq.~(\ref{double_null_metric}) yields
\be
ds^{2}\approx {D^2}(t)\left(-dt^2+dx^2+x^{2}d\Omega^2\right).
\label{conformally_flat}
\ee
So, near the center, the spacetime is nearly conformally flat (NCF).

Some comments on the nearly conformal flatness feature are made as follows:
\begin{enumerate}[fullwidth,itemindent=0em,label=(\roman*)]
\item We also simulate collapse toward BH formation and flat spacetime (dispersion), and find that Eq.~(\ref{eom_r_4}) is also valid in these two cases. So the NCF region also exists near the center in BH formation and dispersion cases.
\item In Ref.~\cite{Bardeen}, the dynamics near the center in axisymmetric system was studied in detail. The local flatness near the center yields the conformal flatness for the spatial part of the metric. This feature and the homogeneity property~\cite{Bardeen} will generate the nearly conformal flatness for the spacetime near the center. For regularization and local flatness at the center, also see Refs.~\cite{Alcubierre:2004gn,Alcubierre_2008}.

  As shown in Fig.~\ref{fig:evolution}(b), near the center, for slice $t=\mbox{Constant}$, there are $\sigma\approx\mbox{Constant}$ and $r(t,x)\propto x$, which should be related to the boundary conditions of $r_{,xx}=\sigma_{,x}\equiv0$ at $x=0$, see Sec.~\ref{sec:framework} and Fig.~\ref{fig:r_xx}. In the Oppenheimer-Snyder spherical dust collapse, the density $\rho$ is uniform, the pressure is zero, and the energy-momentum tensor is $T^{\mu}_{\nu}=\rho(-1,0,0,0)$. The interior spacetime is described by the closed Friedmann-Lema\^{\i}tre-Robertson-Walker metric~\cite{Oppenheimer,Baumgarte_2010}. Obviously, the spacetime is conformally flat. In spherical scalar collapse, as discussed in Sec.~\ref{sec:framework}, near the center, $\psi_{,x}=0$. So near the center, the energy-momentum tensor becomes $T^{\mu}_{\nu}\approx(1/2)e^{2\sigma}\psi_{,t}^2(-1,1,1,1)$ and is almost uniform. Therefore, spherical scalar collapse near the center is similar to the Oppenheimer-Snyder spherical dust collapse, which naturally leads to the nearly conformal flatness feature near the center.
\item The nearly conformal flatness means that the Weyl tensor $C_{\alpha\beta\mu\nu}$, the scalar
$C_{\scriptsize W}\equiv\sqrt{C_{\alpha\beta\mu\nu}C^{\alpha\beta\mu\nu}}$, and the tidal force are nearly zero. Then for the metric~(\ref{double_null_metric}), using Eq.~(\ref{equation_r}), there is
\be
\begin{split}
C_{\scriptsize W}&=\sqrt{\frac{4}{3}}e^{2\sigma}\bigg[-\sigma_{,tt}+\sigma_{,xx}+\frac{-r_{,tt}+r_{,xx}}{r}\\
& \qquad \qquad \quad + \frac{r_{,t}^{2}-r_{,x}^2+e^{-2\sigma}}{r^2}\bigg]\\
&=\sqrt{\frac{4}{3}}e^{2\sigma}\bigg[-\sigma_{,tt}+\sigma_{,xx}+\frac{2(-r_{,tt}+r_{,xx})}{r}\bigg]\\
&\approx 0.
\end{split}
\label{C_w}
\ee
Taking into account Eq.~(\ref{equation_sigma}), one obtains
\be
|-\sigma_{,tt}+\sigma_{,xx}|:\Big|\frac{r_{,tt}-r_{,xx}}{r}\Big|:4\pi|\psi_{,t}^2-\psi_{,x}^2|\approx 2:1:3.
\label{ratios_critical_collapse}
\ee
\item Using $\sigma\approx-\ln{(r/x)}$ and $r(t,x)\approx D(t)x$, one obtains
\be \sigma_{,tt}+\frac{r_{,tt}}{r}-\frac{r_{,t}^2}{r^2}\approx 0, \label{check_1}\ee
\be \sigma_{,xx}+\frac{r_{,xx}}{r}\approx 0. \label{check_2}\ee
Then with $e^{-\sigma}\approx r_{,x}$, the first line of Eq.~(\ref{C_w}) leads to $C_{\scriptsize W}\approx 0$, verifying that the spacetime in which $e^{-\sigma}\approx r/x$ is indeed nearly conformally flat. Moreover, Eqs.~(\ref{check_1}) and (\ref{check_2}) imply that, for the dynamics near the center, the coordinates $t$ and $x$ are not symmetric.
\item We also compute the Weyl tensor in the Roberts solution, which is one type of continuously self-similar critical collapse of a scalar
field~\cite{Roberts,Oshiro:1994hd,Brady:1993np}, and obtain nonzero results. The solution can be written as below~\cite{Oshiro:1994hd}:
\be ds^2=-dudv+r^{2}d\Omega^2,\ee
\be r^2=\frac{1}{4}[(1-k^2)v^2-2vu+u^2],\ee
\be \psi=\pm\frac{1}{2}\ln\frac{(1-k)v-u}{(1+k)v-u},\ee
where $k$ is a parameter. For this solution,
\be
C_{\alpha\beta\mu\nu}C^{\alpha\beta\mu\nu}=\frac{256u^{2}v^{2}k^{4}}{3(-v^2+v^{2}k^2+2vu-u^2)^4},
\ee
which is nonzero for non-Minkowskian spacetime.
\end{enumerate}

\section{Approximate analytic expressions\label{sec:analytic_slns}}
In this section, we present partial analytic information obtained via combination of numerical simulations and asymptotic analysis. In Fig.~\ref{fig:sigma_psi_vs_tau}(a), we plot $-\sigma$ vs. $-\ln\tau$ on the slice of $x=4.75\times 10^{-4}$. $\tau$ is the proper time, defined as $\tau\equiv \int_0^{\xi}e^{-\sigma}d\xi$, where $\xi\equiv t-t_0$ and $t_0$ is the coordinate time when $r$ goes to zero. As shown in Fig.~\ref{fig:sigma_psi_vs_tau}(a), $-\sigma$ is a superposition of a linear function and a periodic function of $-\ln\tau$. In fact, in many discrete scale invariance systems, the log-periodic oscillations exist in the time dependence of the energy release as the impending rupture is approached. The typical time-to-failure formula is~\cite{Sornette_2006,Sornette_1998}
\be E\sim (t_{r}-t)^{n}\left[1+\beta\cos\left(2\pi\frac{\ln(t_{r}-t)}{\ln\lambda}+\varphi\right)\right],\label{log_periodic_1}\ee
where $E$ is the energy released or some other variable describing the on-going damage, $t_{r}$ is the rupture time, $n$ is a critical exponent, $\lambda$ is a preferred scaling factor, and $\varphi$ is a phase. The critical collapse system is also a discrete scale invariance system. So we fit the numerical results of $-\sigma$ vs. $-\ln\tau$ according to the logarithm of Eq.~(\ref{log_periodic_1}),
\be f(\zeta)=a\zeta+\ln[1+b\cos(\omega\zeta+c)]+d. \label{fit_sigma_tau}\ee
Noting that the period for $\sigma$ in Fig.~\ref{fig:sigma_psi_vs_tau}(a) is $T_{\sigma}\approx 1.795$, we fix the angular frequency $\omega$ in Eq.~(\ref{fit_sigma_tau}) as $\omega=2\pi/T_{\sigma}=3.50$. As shown in Fig.~\ref{fig:sigma_psi_vs_tau}(a), the numerical results can be well fit by this formula. The close agreement between the numerical results and the fitting formula strongly implies that critical collapse is indeed one more subfield of complex science.

\begin{figure}[t!]
  \begin{tabular}{ccc}
  \epsfig{file=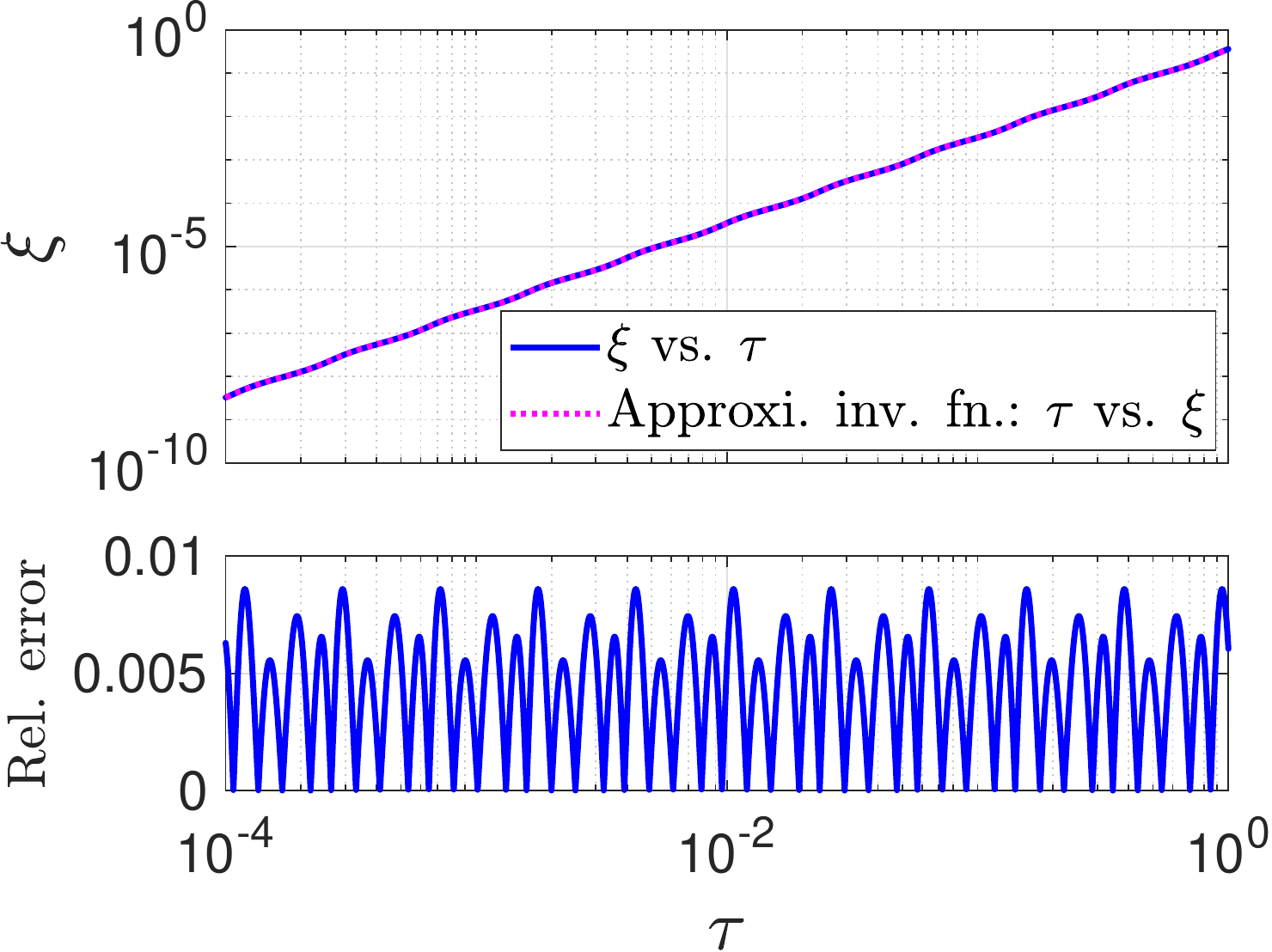,width=8cm}
  \end{tabular}
  \caption{(color online). Upper panel: $\xi$ vs. $\tau$ by Eq.~(\ref{xi_vs_tau}) and the approximate inverse function $\tau$ vs. $\xi$ by Eq.~(\ref{tau_vs_xi}). For both curves in the upper panel, the horizontal axis is for $\tau$, and the vertical one is for $\xi$. Lower panel: The relative error for the two curves in the upper panel.}
  \label{fig:xi_vs_tau}
\end{figure}

\begin{figure*}[t!]
  \begin{tabular}{ccc}
  \epsfig{file=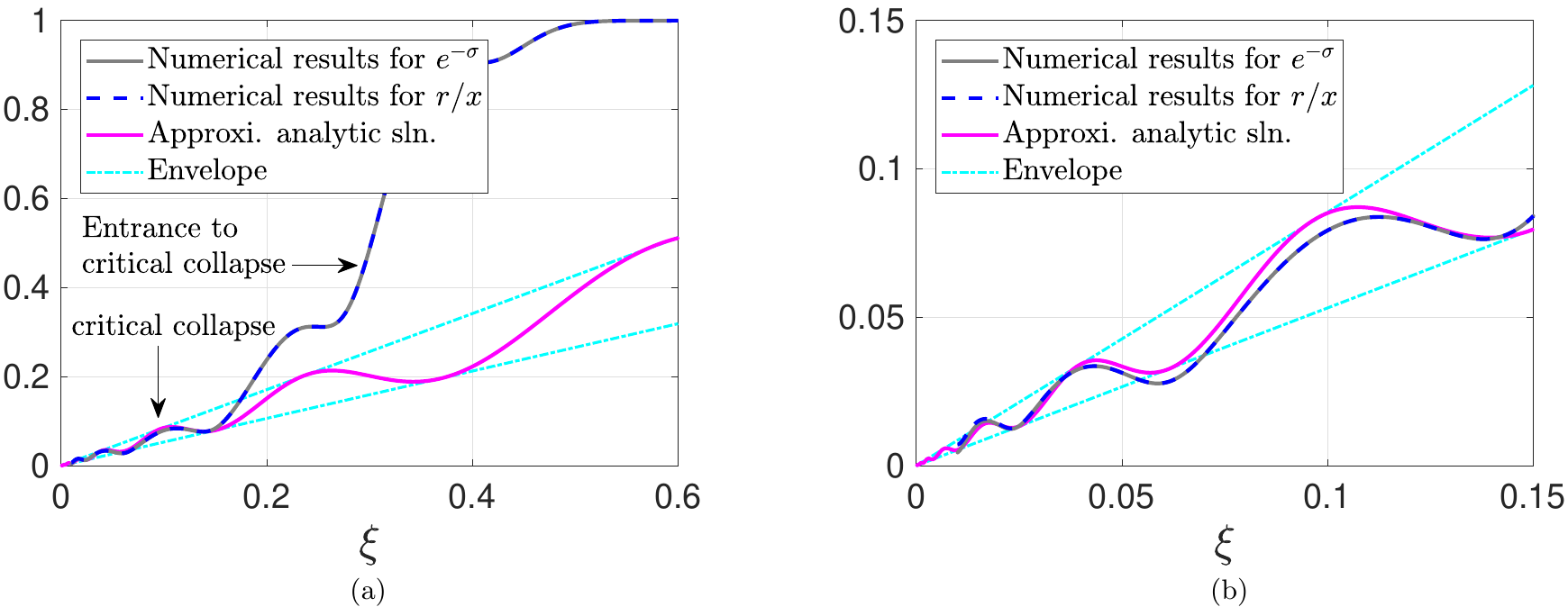, width=0.9\textwidth}
  \end{tabular}
  \caption{(color online). Comparison of the numerical results of $e^{-\sigma}$ and $r/x$ with the approximate analytic expression~(\ref{sigma_analytic}).
  $\xi\equiv t_{0}-t$, and $t_0(=0.6095)$ is the coordinate time when $r$ goes to zero. The regions for $\xi>0.15$ and $\xi<0.15$ correspond to the entrance to critical collapse and formal critical collapse, respectively. Also see Figs.~\ref{fig:eom_r} and \ref{fig:sigma_psi_vs_tau}. Slice: $x=4.75\times 10^{-4}$.}
  \label{fig:sigma_vs_xi}
\end{figure*}

\begin{figure*}[t!]
  \begin{tabular}{ccc}
  \epsfig{file=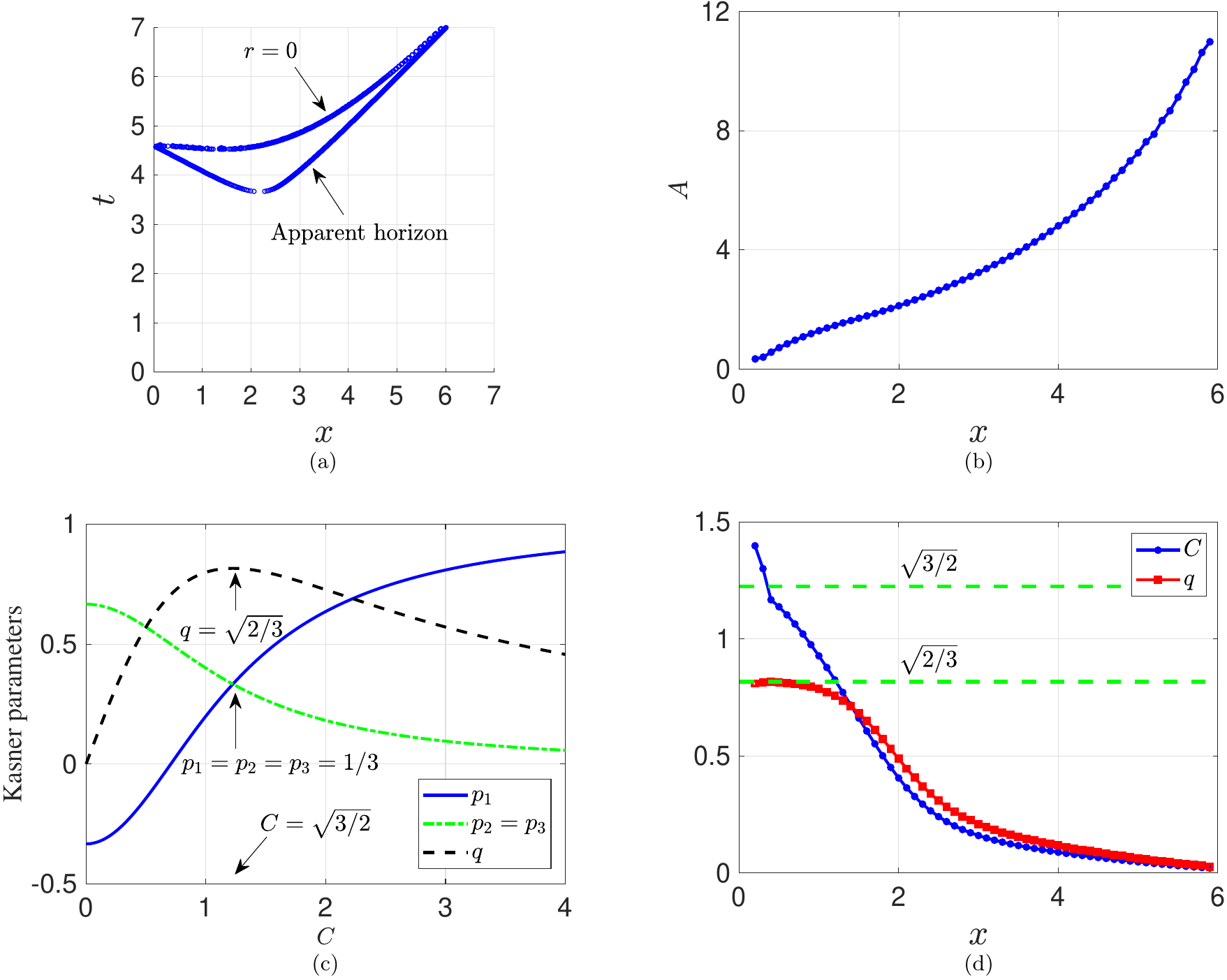, width=0.9\textwidth}
  \end{tabular}
  \caption{(color online). Some numerical results on BH formation presented in Ref.~\cite{Guo:2013dha}.
  (a) Apparent horizon and singularity curve, $r=0$, of the formed BH.
  (b) $A$ for Eq.~(\ref{r_kasner}), $r\approx A\xi^{1/2}$, along the singularity curve. For $x<1$, $A\approx x$.
  (c) The Kasner exponents $p_i$ and the parameter $q$.
  (d) $C$ and $q$ for Eq.~(\ref{psi_kasner}), $\sqrt{8\pi}\psi\approx C\ln\xi\approx q\ln\tau$. At $x=0$ along the curve of $r=0$, $C=\pm\sqrt{3/2}$. This corresponds to a special configuration: (i) A black hole including a central singularity is just formed. (ii) $p_1=p_2=p_3=1/3$, and the metric is asymptotically conformally flat. (iii) $|q|$ approaches its maximum $|q|=\sqrt{2/3}$.}
  \label{fig:BH_formation}
\end{figure*}

With the fitting result~(\ref{fit_sigma_tau}) for $\sigma=\sigma(\tau)$ and $\xi=\int_0^{\tau}e^{\sigma}d\tau$, one can obtain the approximate analytic expression for $\xi=\xi(\tau)$. The fitting result for $a$ in Eq.~(\ref{fit_sigma_tau}) is $-0.5028\pm0.0015$. For simplicity, we set $a=-1/2$. In addition, let $y=\omega\ln\tau-c$ and use Taylor series expansion, there is
\be
\begin{split}
\xi&\approx\frac{e^{-d+c/(2\omega)}}{\omega}\int_{-\infty}^{y}\frac{e^{y/(2\omega)}}{1+b\cos{y}}dy\\
&\approx\frac{e^{-d+c/(2\omega)}}{\omega}\int_{-\infty}^{y}{e^{y/(2\omega)}}(1-b\cos{y}+b^{2}\cos^{2}y)dy\\
&\approx e^{-d}(2+b^2)\tau^{1/2}\left[1-H\sin(\omega\ln\tau-c+\delta)\right],
\end{split}
\label{xi_vs_tau}
\ee
where $H=2b/[(2+b^2)\sqrt{1+4\omega^2}]\approx0.03$, and $\delta=\arccos[2\omega/\sqrt{1+4\omega^2}]\approx 0.14$ radians. Noting that $H\approx 0.03\ll 1$, the major part for the inverse function $\tau=\tau(\xi)$ is
\be \tau_m=\left(\frac{e^{d}\xi}{2+b^2}\right)^2.\label{tau_vs_xi_major}\ee
Substitution of Eq.~(\ref{tau_vs_xi_major}) into the second term in the square brackets in Eq.~(\ref{xi_vs_tau}) yields an approximate analytic expression for $\tau=\tau(\xi)$
\be
\tau^{1/2}\approx\frac{e^{d}\xi}{2+b^2}\left[1+H\sin(2\omega\ln\xi+\eta+\delta)\right],
\label{tau_vs_xi}
\ee
where $\eta=2\omega[d-\ln(2+b^2)]-c\approx-2.32$. The relative error for Eq.~(\ref{tau_vs_xi}) compared to Eq.~(\ref{xi_vs_tau}) is less than $1\%$. See Fig.~\ref{fig:xi_vs_tau}.

Combining Eqs.~(\ref{eom_r_4}), (\ref{fit_sigma_tau}), and (\ref{tau_vs_xi}), one obtains the first approximate analytic expression for the metric components near the center
\be
\begin{split}
e^{-\sigma}&\approx\frac{r}{x}\\
&\approx{e^{d}}\tau^{1/2}[1+b\cos(\omega\ln\tau-c)]\\
&\approx\frac{e^{2d}\xi}{2+b^2}\Big\{1+b\cos(2\omega\ln\xi+\eta)\\
&\qquad \qquad \quad \ +H\sin(2\omega\ln\xi+\eta+\delta)\\
&\qquad \qquad \quad \ +\frac{Hb}{2}[\sin(4\omega\ln\xi+2\eta+\delta)+\sin\delta]\Big\}.
\end{split}
\label{sigma_analytic}
\ee
As shown in Fig.~\ref{fig:sigma_vs_xi}, Eq.~(\ref{sigma_analytic}) matches well with the numerical results for $\xi<0.15$, which corresponds to formal critical collapse. There is a large deviation between the analytic expression~(\ref{sigma_analytic}) and the numerical results for $\xi>0.15$, which should be because that this region is not a formal critical collapse region, but an entrance to critical collapse. Also see Figs.~\ref{fig:eom_r} and \ref{fig:sigma_psi_vs_tau}.

The numerical results on $\psi$ vs. $-\ln\tau$ are plotted in Fig.~\ref{fig:sigma_psi_vs_tau}(b). Roughly speaking, $\psi$ is a linear function of $\ln\tau$ and switches between $\alpha\ln\tau$ and $-\alpha\ln\tau$ after every half-period with $\alpha\approx: 0.6\sim0.8$. The numerical results on $\psi$ vs. $-\ln\tau$ are fit according to fourth-order Fourier sine and cosine series. The fitting results are shown in the caption of Fig.~\ref{fig:sigma_psi_vs_tau}. In Sec.~\ref{sec:comparison}, we will give one preliminary analytic explanation to the slopes of $(-\sigma,\psi)$ vs. $-\ln\tau$.

\section{Critical collapse vs. dynamics near spacetime singularities\label{sec:comparison}}
A naked singularity is formed in critical collapse. Therefore, it is meaningful to connect critical collapse to the results on the dynamics near spacetime singularities that have been obtained before, including the BKL conjecture and BH formation. We will do so in this section.

\subsection{BH formation by spherical scalar collapse}
The BKL conjecture is one of the guiding principles in the studies of spacetime singularities. According to the BKL conjecture, the behaviors of the spacetime and the matter field are described by the Kasner solution~\cite{Belinskii,Landau,Nariai,Belinskii_2,Kasner,Wainwright,Kamenshchik:2010jz,Belinski:2014kba,Scheel:2014hfa}. The four-dimensional homogeneous but anisotropic Kasner solution with a massless scalar field $\psi$ minimally coupled to gravity can be written as follows~\cite{Nariai,Kamenshchik:2010jz,Belinski:2014kba}:
\be
ds^2=-d\tau^2+\sum\limits_{i=1}^3 \tau^{2p_i}dx_{i}^{2}.
\label{Kasner_solution}
\ee
The Kasner exponents satisfy $p_1+p_2+p_3=1$ and $p^{2}_{1}+p^{2}_{2}+p^{2}_{3}=1-q^2$. The parameter $q$ describes the contribution of the scalar field, $\sqrt{8\pi}\psi=q\ln\tau$. For reviews on the BKL conjecture, see Refs.~\cite{Kamenshchik:2010jz,Belinski:2014kba,Scheel:2014hfa}. In BH formation by spherical scalar collapse, the dynamics near the central singularity is described by the Kasner solution~\cite{Guo:2013dha}. On slice of $x=\mbox{Constant}$, there are
\be r\approx A\xi^{\frac{1}{2}}\approx A\left(\frac{3+2C^2}{4}\tau\right)^{\frac{2}{3+2C^2}}, \label{r_kasner}\ee
\be e^{-\sigma}\approx\xi^{\frac{-1+2C^2}{4}}\approx\left(\frac{3+2C^2}{4}\tau\right)^{\frac{-1+2C^2}{3+2C^2}},\label{sigma_kasner}\ee
\be \sqrt{8\pi}\psi\approx C\ln\xi\approx\frac{4C}{3+2C^2}\ln\tau,\label{psi_kasner}\ee
where $\tau\equiv\int_0^{\xi}e^{-\sigma}d\xi$, $\xi\equiv t_0-t$, and $t_0$ is the coordinate time on the singularity curve $r=0$ shown in Fig.~\ref{fig:BH_formation}(a). Comparing Eqs.~(\ref{r_kasner})-(\ref{psi_kasner}) to (\ref{Kasner_solution}), we extract $p_1=(-1+2C^2)/(3+2C^2)$, $p_2=p_3=2/(3+2C^2)$, and $q=4C/(3+2C^2)$. In critical collapse, $e^{-\sigma}\approx r/x$. We check whether this is also true in BH formation. Enforcing $e^{-\sigma}\sim r$, one obtains from Eqs.~(\ref{r_kasner}) and (\ref{sigma_kasner})
\be C=\pm\sqrt{\frac{3}{2}}.\label{D_kasner}\ee
As shown in Fig.~\ref{fig:BH_formation}(a) and \ref{fig:BH_formation}(d), the result of $C=\pm\sqrt{3/2}$ corresponds to the spacetime point of $x=0$ along the singularity curve $r=0$, where the center is just transiting from regular into singular. As shown in Fig.~\ref{fig:BH_formation}(c), in this configuration, $p_1=p_2=p_3=1/3$, so the metric is asymptotically conformally flat. Actually, as shown in Fig.~\ref{fig:BH_formation}(b), for $x<1$, the `coefficient' $A$ in Eq.~(\ref{r_kasner}) has a rough relation $A\approx x$. Therefore, similar to critical collapse, in BH formation, for $x$ close to zero along the singularity curve, there is also $e^{-\sigma}\approx{r/x}$.

In the case of $C=\pm\sqrt{3/2}$, Eqs.~(\ref{sigma_kasner}) and (\ref{psi_kasner}) can be rewritten as
\be \sigma\approx-\frac{1}{3}\ln\tau,\label{sigma_tau_BH}\ee
\be \sqrt{8\pi}\psi\approx q\ln\tau\approx\pm\sqrt{\frac{2}{3}}\ln\tau.\label{psi_tau_BH}\ee
In the metric~(\ref{Kasner_solution}), the Kreschmann scalar is~\cite{Guo:2015ssa}
\be
K=\frac{12J}{(3+2C^2)^4}\tau^{-4}\approx\frac{3J}{64}\left(\frac{r}{A}\right)^{-2(3+2C^2)}.
\label{K_scalar_BH_formation}
\ee
where $J=16-40C^2+99C^4-26C^6+3C^8$. In the case of $C=\pm\sqrt{3/2}$, there is $K\propto r^{-12}$.

\subsection{Critical collapse vs. BH formation}
In Table~\ref{tab:Table1_compare}, we compare the dynamics of critical collapse near the center and that at $x=0$ along the singularity curve in BH formation. Some discussions are given below.

In BH formation, the scalar field and gravity are strong. Then, the scalar field is simply absorbed into the central singularity without reflections, $\psi=\alpha\ln\tau$ or $\psi=-\alpha\ln\tau$ with $\alpha\approx0.16$, and $\sigma=-(1/3)\ln\tau$. In critical collapse, the scalar field and gravity are weak, and there is a balance between gravity and reflections at the center. Consequently, $\psi$ switches between $\alpha\ln\tau$ and $-\alpha\ln\tau$ with $\alpha\approx:0.6\sim0.8$, and $\sigma$ is a superposition of a linear function and a periodic function of $\ln\tau$ with the slope for the linear part being about $-1/2$.

The slopes of $(\sigma,\psi)$ vs. $\ln\tau$ in critical collapse are at the same order of magnitude as those in BH formation, respectively. However, as shown in Eqs.~(\ref{sigma_tau_BH}) and (\ref{psi_tau_BH}), the values of the slopes in BH formation can be obtained analytically. In this sense, for the first time, we give one rough analytic explanation to the slopes of $(\sigma,\psi)$ vs. $\ln\tau$ in critical collapse.

The presence of the cosine function in the second line of Eq.~(\ref{sigma_analytic}) implies that the dynamics in critical collapse is not described by the  Kasner solution. Therefore, the BKL conjecture is not valid for the naked singularity formed in critical collapse, which is different from BH singularities.

\begin{figure}[t!]
  \begin{tabular}{ccc}
  \epsfig{file=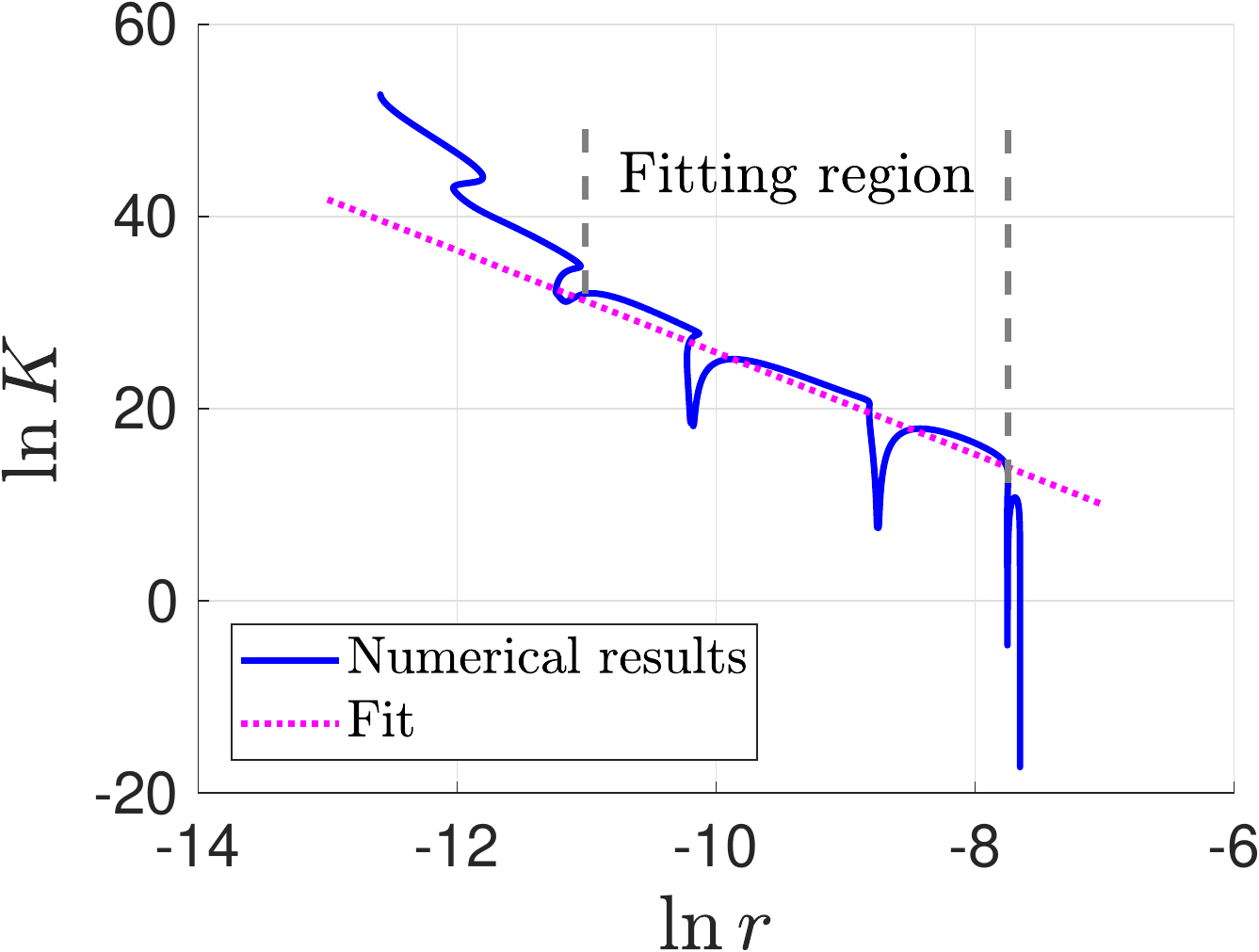,width=7.3cm}
  \end{tabular}
  \caption{(color online). $\ln{K}$ vs. $\ln{r}$, where $K$ is the Kreschmann scalar. The relation of $\ln{K}$ vs. $\ln{r}$ is fit according to $f(x)=ax+b$ with $a=-5.30\pm0.04$ and $b=-27.16\pm0.38$. Slice: $x=4.75\times 10^{-4}$.}
  \label{fig:K vs_r}
\end{figure}

As shown in Eq.~(\ref{K_scalar_BH_formation}), in BH formation, in the case of $C=\pm\sqrt{3/2}$, the Kreschmann scalar $K\propto r^{-12}$. We plot the Kreschmann scalar in critical collapse in Fig.~\ref{fig:K vs_r}, and observe that $K\propto r^{-5.30}$, much weaker than the BH formation case, and also weaker than the Schwarzschild BH case where $K\propto r^{-6}$.

In summary, critical collapse and BH formation share some common features: after all they both are on the dynamics in the vicinities of singularities. On the other hand, there are some remarkable differences between the two: the scalar field and gravity are weaker in critical collapse than in BH formation.

\begin{table*}
\centering
\caption{BH formation vs. critical collapse}
\renewcommand{\arraystretch}{1.3}
\newcommand{\tabincell}[2]{\begin{tabular}{@{}#1@{}}#2\end{tabular}}
\begin{tabular}{c|c}
\Xhline{0.8pt}
{\bf \tabincell{c}{$x=0$ on the singularity curve in BH formation}}&{\bf NCF region in critical collapse}\\
\Xhline{0.8pt}
\tabincell{c}{From beginning of collapse to formation of singularity:\\
 $e^{-\sigma}\approx \frac{r}{x}$}
 &\tabincell{c}{$e^{-\sigma}\approx \frac{r}{x}$}\\
\hline
$\sigma\approx-\frac{1}{3}\ln\tau$&$\sigma\approx-\frac{1}{2}\ln\tau-\ln[1+0.23\cos(3.50\ln\tau+1.48)]-0.18$\\
\hline
\tabincell{c}{$\psi$ is absorbed into the central singularity, \\ and $\psi=\alpha\ln\tau$ or $-\alpha\ln\tau$ with $\alpha\approx\sqrt{1/12\pi}\approx0.16$.}
  &\tabincell{c}{$\psi$ is reflected at $x=0$, and switches between $\alpha\ln\tau$\\and $-\alpha\ln\tau$ with $\alpha\approx:0.6\sim0.8$ after each half-period.}\\
\hline
The spacetime is nearly conformally flat.&The spacetime is nearly conformally flat.\\
\hline
The BKL conjecture applies.&The BKL conjecture does not apply.\\
\hline
Kreschmann scalar $\propto r^{-12}$ & Kreschmann scalar $\propto r^{-5.30}$\\
\Xhline{0.8pt}
\end{tabular}
\label{tab:Table1_compare}
\end{table*}

\section{Scalar field in critical collapse vs. standing waves\label{sec:standing_wave}}
In critical collapse, the scalar field moves toward the center under gravity. At the same time, the scalar field is reflected at the center. The resulting scalar field is discretely self-similar. This picture makes us to speculate that the scalar field in critical collapse may be a special standing wave. The word `special' is due to the fact that, in critical collapse, the scalar wave keeps shrinking toward the center. We think that this special standing wave picture helps to interpret the DSS feature and to seek the full analytic solution to critical collapse.

In fact, standing waves are quite common in astrophysics: e.g., (i) The 5-minute solar oscillation in local surface velocities discovered by Leighton~\cite{Neugebauer_1997,Li_2012}. (ii) The density wave theory on the spiral structure of disk galaxies by Lin and Shu~\cite{Lin_1964,Lin_1966,Lin_1967,Yuan_2006,Xiang_2008}. (iii) The microwave cavity hypothesis on ball lighting by Kapitsa~\cite{Kapitsa_1955a,Kapitsa_1955b}.

\section{Discussions\label{sec:discussions}}
We obtained the first approximate analytic expression for the metric near the center in critical collapse. We observed that in spherical scalar collapse, the spacetime near the center is nearly conformally flat. The interplay between critical collapse and complex science deserves to be explored further, which is meaningful for both sides. In critical collapse, gravity and reflections at the center compete and compromise. Similar mechanisms widely exist in other branches of complex science~\cite{Li_2014}. The BKL conjecture may still be a guiding principle to achieve deeper understanding on critical collapse.

Critical collapse is an interdisciplinary subject, connecting gravitation (including BH physics and spacetime singularities), complex science, and partial differential equations, etc. Studies in critical collapse can enrich each of these, and deserve further efforts.

\section*{Acknowledgments}
The authors are grateful to Zhoujian Cao, Carsten Gundlach, David Hilditch, Li-Wei Ji, Pankaj S. Joshi, Prashant Kocherlakota, Yun-Kau Lau, Junbin Li, Daoyan Wang, and Xuefeng Zhang for useful discussions, and Beijing Normal University and Sun Yat-sen University for hospitality. This work is supported by the National Natural Science Foundation of China (NSFC) under grant No.11575083, and Shandong Province Natural Science Foundation under grant No.ZR2018MA046 and No.ZR2019MA068.

\appendix*
\section{Derivations of $m_{,t}$ and $m_{,x}$\label{sec:appendix_derivations}}
Taking the first-order temporal derivative on Eq.~(\ref{mass_definition}), one obtains
\be
-\frac{m_{,t}}{r}+\frac{mr_{,t}}{r^2}=e^{2\sigma}[-r_{,t}(r_{,t}\sigma_{,t}+r_{,tt})+r_{,x}(r_{,x}\sigma_{,t}+r_{,xt})].
\label{m_definition_dt}
\ee
Subtraction of Eq.~(\ref{equation_r_2}) from Eq.~(\ref{constraint_eq_xx_tt}) yields
\be
r_{,t}\sigma_{,t}+r_{,tt}=-2\pi r(\psi_{,t}^2+\psi_{,x}^2)-e^{-2\sigma}\frac{m}{r^2}-r_{,x}\sigma_{,x}.
\label{r_sigma_dt}
\ee
Rewrite the constraint equation~(\ref{constraint_eq_xt}) as
\be r_{,x}\sigma_{,t}+r_{,tx}=-4\pi r\psi_{,t}\psi_{,x}-r_{,t}\sigma_{,x}.
\label{constraint_eq_xt_2}
\ee
Substitutions of Eqs.~(\ref{r_sigma_dt}) and (\ref{constraint_eq_xt_2}) into (\ref{m_definition_dt}) yield Eq.~(\ref{dmdt}):
\be
\begin{split}
m_{,t}&=4\pi r^2(r_{,t}T^{t}_{t}-r_{,x}T^{t}_{x})\\
&=4\pi r^2\cdot e^{2\sigma}\left[-\frac{1}{2}r_{,t}(\psi_{,t}^2+\psi_{,x}^2)+r_{,x}\psi_{,t}\psi_{,x}\right].
\end{split}
\ee

Taking the first-order spatial derivative on Eq.~(\ref{mass_definition}), one obtains
\be
-\frac{m_{,x}}{r}+\frac{mr_{,x}}{r^2}=e^{2\sigma}[-r_{,t}(r_{,t}\sigma_{,x}+r_{,xt})+r_{,x}(r_{,x}\sigma_{,x}+r_{,xx})].
\label{m_definition_dx}
\ee
Addition of Eqs.~(\ref{constraint_eq_xx_tt}) and (\ref{equation_r_2}) yields
\be
r_{,x}\sigma_{,x}+r_{,xx}=-2\pi r(\psi_{,t}^2+\psi_{,x}^2)+e^{-2\sigma}\frac{m}{r^2}-r_{,t}\sigma_{,t}.
\label{r_sigma_dx}
\ee
Rewrite the constraint equation (\ref{constraint_eq_xt}) as
\be r_{,t}\sigma_{,x}+r_{,tx}=-4\pi r\psi_{,t}\psi_{,x}-r_{,x}\sigma_{,t}.
\label{constraint_eq_xt_3}
\ee
Substitutions of Eqs.~(\ref{r_sigma_dx}) and (\ref{constraint_eq_xt_3}) into (\ref{m_definition_dx}) yield Eq.~(\ref{dmdx}):
\be
\begin{split}
m_{,x}&=4\pi r^2(r_{,x}T^{x}_{x}-r_{,t}T^{x}_{t})\\
&=4\pi r^2\cdot e^{2\sigma}\left[\frac{1}{2}r_{,x}(\psi_{,t}^2+\psi_{,x}^2)-r_{,t}\psi_{,t}\psi_{,x}\right].
\end{split}
\ee
%


\end{document}